\newcommand{\be}{\begin{equation}}
\newcommand{\ee}{\end{equation}}
\newcommand{\br}{\begin{eqnarray}}
\newcommand{\er}{\end{eqnarray}}

\documentstyle[aps,amssymb,psfig,color,multicol]{revtex}

\begin{document}

\title{Characterization and quantification of symmetric Gaussian
state entanglement through a local classicality criterion}
\author{Marcos C. de Oliveira\footnote{marcos@ifi.unicamp.br}}
\address{
 Instituto de F\'\i sica Gleb Wataghin,  Universidade Estadual de Campinas, 13083-970, Campinas - SP, Brazil.}
\date{\today}

\maketitle
 \begin{abstract}
A necessary and sufficient condition for characterization and
quantification of entanglement of any bipartite Gaussian state
belonging to a special symmetry class is given in terms of
classicality measures of one-party states. For Gaussian states
whose local covariance matrices have equal determinants it is
shown that separability of a two-party state and classicality of
one party state are completely equivalent to each other under a
nonlocal operation, allowing entanglement features to be
understood in terms of any available classicality measure.

\end{abstract}
\begin{multicols}{2}
\narrowtext \tighten

\section{Introduction}

Reliable implementations of quantum communication protocols are
constrained by the actual ability to manipulate quantum systems to
encode both qubits (superpositions of mutually orthogonal states)
and quantum channels (non-locally multiparty entangled states) of information. 
 While pure qubits can be prepared under certain conditions on local
operations, ``good'' (pure) quantum channels are hardly achieved
due to quantum noise. Efforts have been devoted to both classify
and to quantify entanglement in multipartite quantum systems
\cite{vedral}, and to quantify the limiting bounds for efficiency
of many communication protocols against mixing \cite{effprot}.
While entanglement for qubits systems is nowadays quite well
understood, the existent quantifications of entanglement can
hardly be calculated, even numerically, for continuous variable
states (see \cite{cirarcentmeas} and references therein). It is of
central importance thus to develop new scenarios of reasoning on
entanglement
in terms of accessible measures.
  Recent efforts in that direction include the  Horodecki and Ekert \cite{hodekert} direct detection scheme
 for bipartite qubit states; the Kim {\it et al.}
\cite{kimmunro} proposal in terms of linear operations and
homodyning detection for Gaussian continuous variable quantum
states;
 and the P. Marian {\it et al.}
\cite{paulina2} proposal in terms of Bures \cite{bures} distance
measures of bipartite Gaussian states.

 In this paper we outline an alternative approach to understand and quantify
entanglement for Gaussian bipartite input quantum states in terms
of one-party output states nonclassicality aspects. By input and
output we refer to any bosonic system before and after,
respectively, a bilinear Bogoliubov (beam-splitter similar)
operation. This approach extends the applicability of a recent
theorem providing that if the input state is classical, the output
state of a beam splitter must be a separable state \cite{buzek}.
Remarkably, the inverse problem sets a necessary and sufficient
condition for entanglement quantification of a special symmetry
class of Gaussian states if no other local operations are made on
the output ports. Since under the specified operation Gaussian
measures for the bipartite input state are transformed into output
one-party Gaussian measures, the degree of entanglement of the
input can be quantified in terms of any available nonclassicality
measure of the Gaussian type for one (or both) of  the output
ports.
 We address to
the identification of the negativity of the output Glauber
$P$-function \cite{englert}
for classicality check, and exemplify the above relation by
quantifying the
 entanglement of a two-mode thermal squeezed states in terms
 of the output one-party nonclassicality measure - the Bures distance
 from a one-mode squeezed state \cite{paulina2,paulina1}.

 We begin  by reviewing in Sec. II some properties of bipartite
Gaussian states, including the necessary and sufficient conditions
for the state to be separable. In Sec. III we analyze the
classicality of an arbitrary bipartite Gaussian state under a
nonlocal bilinear Bogoliubov operation. In Sec. IV we present the
way to quantify entanglement in terms of the transformed output
one-port classicality measure. In Sec. V we exemplify our
discussion by quantifying the amount of entanglement  of a
two-mode thermal squeezed states by means of the output Bures
distance from an one-mode squeezed state. Finally in Sec. VI a
conclusion encloses the paper.

\section{Bipartite Gaussian States}

 A
bipartite quantum state $\rho$ is Gaussian if its symmetric
characteristic function is given by 
$C({\bbox{\eta}})=Tr[D({\bbox{\eta}})\rho]=e^{-\frac12{\bbox{\eta}^\dagger}{\bf
V}{\bbox{\eta}} }$, where
$D(\bbox{\eta})=e^{-\bbox{\eta}^\dagger{\bf E}{\bf v}}$ is a
displacement operator in the parameter four-vector
$\bbox{\eta}$-space: 
$\bbox{\eta}^\dagger=\left(\eta_1^*, \eta_1, \eta_2^*,
\eta_2\right)$, ${\bf v}^\dagger= \left(a_1^\dagger, a_1,
a_2^\dagger, a_2\right)$, and
  \be {\bf
E}=\left(\begin{array}{c c}{\bbox{Z}}&{\bf 0}\\{\bf 0}&
{\bbox{Z}}\end{array}\right),\;\;\;
{\bbox{Z}}=\left(\begin{array}{c c}{1}&{0}\\{0}&
{-1}\end{array}\right),\ee where $a_1$ ($a_1^\dagger$) and $a_2$
($a_2^\dagger$) are annihilation (creation) operators for party 1
and 2, respectively. ${\bf V}$ is a $4\times 4$ covariance matrix
with elements
$V_{ij}=(-1)^{i+j}\langle\{v_i,v_j^\dagger\}\rangle/2$, which can
be decomposed in four block $2\times 2$ matrices, \be\label{var}
{\bf V}=\left(\begin{array}{c c}{\bf V_1}&{\bf C}\\{\bf
C}^\dagger& {\bf V_2}\end{array}\right)=\left(\begin{array}{c c
c c}n_1&m_1&m_s&m_c\\m^*_1& n_1&m^*_c&m^*_s\\
m^*_s&m_c&n_2&m_2\\ m_c^*&m_s&m_2^*&n_2\end{array}\right),\ee
where ${\bf V_1}$ and ${\bf V_2}$ are Hermitian matrices
containing only local elements while ${\bf C}$ is the correlation
between the two parties. Positivity and separability for bipartite
Gaussian quantum states have been largely investigated
\cite{englert,simon,schur}.
 Besides
the requirement of positivity  \be \label{cond}{\bf V}+\frac 12
{\bf E}\ge 0,\ee which is nothing but the fundamental uncertainty
principle,
  a necessary and sufficient condition for separability of
Gaussian states is that they must also follow \cite{schur} \be
{\bf \widetilde V}+\frac 12 {\bf E}\ge 0,\label{cond2}\ee under a
partial phase space mirror
reflection 
 ${\bf\widetilde V}={\bf TVT}:{\bf T v^\dagger}={\bf v^\dagger}_T=
 \left(\begin{array}{c c c c}a_1^\dagger&
a_1&a_2&a_2^\dagger\end{array}\right)$, with \be {\bf
T}=\left(\begin{array}{c c}{\bf I}&{\bf 0}\\{\bf 0}& {\bf
X}\end{array}\right),\, \text{and}\, {\bf X}=\left(\begin{array}{c
c}{ 0}&{ 1}\\{1}& {0}\end{array}\right).\ee

Through Schur block decomposition \cite{schur,higham} the physical
positivity criterion (\ref{cond}) applies only if \be {\bf
V_1}+\frac 12 {\bf Z}\ge 0,\;\;\left({\bf V_2}+\frac 12 {\bf
Z}\right)-{\bf C}^\dagger\left({\bf V_1}+\frac 12 {\bf
Z}\right)^{-1}{\bf C}\ge 0,\ee which explicitly reads
 \br \label{sep}n_1&\ge&
\sqrt{|m_1|^2+\frac 1 4},\\ n_2&\ge& \frac{s}{d}+\sqrt{\frac1 4
\left[\frac{\left||m_c|^2-|m_s|^2\right|}{d}-1\right]^2+|m_2-c|^2},\er
respectively,  with $
s=n_1\left(|m_c|^2+|m_s|^2\right)-m_cm_sm_1^*-m_c^*m_s^*m_1$,
$c=2n_1m_s^*m_c-m_c^2m_1^*-(m_s^*)^2m_1$, and $ d=n_1^2-\frac1
4-|m_1|^2$. Similarly
 the separability condition (\ref{cond2}) writes explicitly into (\ref{sep}) and
 \br \label{sep2}n_2&\ge& \frac{s}{d}+\sqrt{\frac1 4
\left[\frac{\left||m_c|^2-|m_s|^2\right|}{d}+1\right]^2+|m_2-c|^2}.\er
%

\section{Classicality of the output state}
 The above general criterion over the two party state observables
can be translated into directly measurable quantities of one party
under nonlocal operations. 
Let a Gaussian entangled state $\rho_{in}$ be transformed under a
nonlocal bilinear Bogoliubov operation $B$:
\begin{eqnarray}
\rho_{out}&=&B\rho_{in}B^\dagger,\label{1}\\
B{\bf v}B^\dagger
&=&{\bf M} {\bf v},\label{2}\\
{\bf M}&=&\left(\begin{array}{c c}{\bf R}&{\bf S}\\{\bf -S}^*&
{\bf R}^*\end{array}\right)\\
 {\bf R}=\cos\theta\left(\begin{array}{cc} e^{i\phi_0}&0\\0&
e^{-i\phi_0}\end{array}\right),&{\bf
S}&=\sin\theta\left(\begin{array}{cc}e^{i\phi_1}&0\\0&e^{-i\phi_1}\end{array}\right)
\end{eqnarray} where
$\rho_{out}$ is the density operator for the joint output state.
The output symmetric characteristic function is given by \be
C_{out}({\bbox{\eta}})=Tr[D({\bbox{\eta}})\rho_{out}]=Tr[B^\dagger
D({\bbox{\eta}})B\rho_{in}].\ee Now with the help of Eqs.(11-13),
$ B^\dagger D({\bbox{\eta}})B=e^{-\bbox{\eta}^\dagger{\bf E}{\bf
M}^{-1}{\bf v}}\equiv D({\bbox{\zeta}})$, with $\bbox{\zeta}={\bf
M}\bbox{\eta}$, since ${\bf M} {\bf E M}^{-1}={\bf E}$. Thus \be
C_{out}({\bbox{\eta}})=C_{in}({\bbox{\zeta}})=e^{-\frac12{\bbox{\zeta}^{\dagger}}{\bf
V}{\bbox{\zeta}} }=e^{-\frac12{\bbox{\eta}^{\dagger}}{\bf
V^\prime}{\bbox{\eta}} },\label{car}\ee
 where ${\bf
V}^\prime={\bf M}^{-1}{\bf V}{\bf M}$, and 
analogously to (\ref{var}), ${\bf V}^\prime$ can be block
decomposed
with \br {\bf V_1^\prime}&=& {\bf R}^{*}{\bf V_1}{\bf R}+{\bf
S}{\bf V_2}{\bf S}^*-{\bf
S}{\bf C}^\dagger{\bf R}-{\bf R}^{*}{\bf C}{\bf S}^{*},\\
 {\bf V_2^\prime}&=& {\bf
S}^{*}{\bf V_1}{\bf S}+{\bf R}{\bf V_2}{\bf R}^{*}+{\bf R}{\bf
C}^\dagger{\bf S}+{\bf S}^{*}{\bf C}{\bf R}^{*},\\
 {\bf C^\prime}&=& {\bf
R}^{*}{\bf V_1}{\bf S}-{\bf S}{\bf V_2}{\bf R}^{*}-{\bf S}{\bf
C}^\dagger{\bf S}+{\bf R}^{*}{\bf C}{\bf R}^{*},\er allowing the
following important results.

 The output state is separable, such that $
C_{out}({\bbox{\eta}})=C_{out}(\bbox{\eta_1})C_{out}(\bbox{\eta_2})$,
if and only if ${\bf C^\prime}={\bf 0}$, which explicitly means
\br \label{c1}\sin 2\theta&&\left(m_2e^{i(\phi_0+\phi_1)}-m_1
e^{-i(\phi_0+\phi_1)}\right)-2\cos2\theta m_c=0,\\
\sin2\theta&& e^{-i(\phi_0-\phi_1)}(n_1-n_2)+\nonumber\cos2\theta
\left(m_s e^{-2i\phi_0}+m_s^* e^{2i\phi_1}\right)\\&&+\left(m_s
e^{-2i\phi_0}-m_s^* e^{2i\phi_1}\right)=0.\label{c2}\er
 From now on we fix $\theta=\pi/4$, setting the equivalence of the transformation
  ${\bf M} $ to that of an ideal 50:50 beam-splitter. Thus
  conditions (\ref{c1}) and (\ref{c2}) reduce to
\br \label{c1l}&&m_2e^{i(\phi_0+\phi_1)}-m_1
e^{-i(\phi_0+\phi_1)}=0,\\
&& e^{-i(\phi_0-\phi_1)}(n_1-n_2)+\left(m_s e^{-2i\phi_0}-m_s^*
e^{2i\phi_1}\right)=0.\label{c2l}\er Although restrictive those
conditions can be naturally reached for special classes of
Gaussian states as we remark bellow.

 {\bf Remark
1:} Important examples of bipartite Gaussian states existent in
Nature have a symmetric special form $n_1=n_2$, and
$m_1=m_2=m_s=0$
 \cite{simon,schur} that satisfy
 conditions (\ref{c1l}) and (\ref{c2l}) automatically. Such is the case for the two-mode
thermal squeezed state (TMTSS) \cite{daffer} generated in a
nonlinear crystal with internal noise. There $n_1=n_2=n=g h_1$,
$m_1=m_2=m_s=0$, and $m_c=m=g h_2$, with $g=h_1h_2/(h_1^2-h_2^2)$
and $
 h_i=\left[e^{-p_i}+d\,(2\bar n
+1)\left({(1-e^{-p_i})}/{p_i}\right)\right]$, where $p_1=d+2r$,
and $p_2=d-2r$. $d=\gamma t$ is a diffusion parameter with the
relaxation constant $\gamma$, and $r=\kappa t$ is the squeezing
parameter. $\bar n$ is the mean number of thermal photons
introduced by the quantum noise. Latter on we consider this
specific situation.

{\bf Remark 2:} Under a local $Sp(2,R)\otimes Sp(2,R)$ operation
(see \cite{simon,schur}) an arbitrary covariance matrix
$\mathbf{V}$ can be transformed to the special form of Remark 1,
without affecting the entanglement if, previously to the local
operation, $\mathbf{V}$ is as such that the condition
$n_1^2-|m_1|^2=n_2^2-|m_2|^2$ applies. Those are Gaussian states
whose local covariance matrices,$\mathbf{V_1}$ and $\mathbf{V_2}$
have the same determinant, {\it i.e.},
$\det{\mathbf{V_1}}=\det{\mathbf{V_2}}$. We shall call this subset
of Gaussian states the Set of Gaussian states with Symmetric Local
Determinants (SSLD).
Thus (\ref{c1l}) and (\ref{c2l}) restrict the generality of our
results to the class of Gaussian states belonging to the SSLD,
satisfying $n_1^2-|m_1|^2=n_2^2-|m_2|^2$. Naturally this last
condition is less restrictive than the one required in (\ref{c2l})
and offers a wider range of states that can be covered than those
of the special symmetric form of Remark 1. Hereafter we consider
only this class of Gaussian states.

Under that condition, both the output state positivity and
separability criteria on ${\bf V^\prime}$ give equivalently ${\bf
V^\prime_1 }+\frac 12 {\bf Z}\ge
 0$ and ${\bf V^\prime_2}+\frac 12 {\bf Z}\ge
 0$, or explicitly
 \br \label{pos} n_1^\prime \ge \sqrt{|m_1^\prime|^2+\frac 1 4},\;\;\;
 n_2^\prime \ge \sqrt{|m_2^\prime|^2+\frac 1
4},\er in terms of the primed coefficients of the output variance
matrix, related to the input coefficients by Eqs. (16-18). By
writing the inequalities (\ref{pos}) explicitly in terms of the
input coefficients we obtain the physical positivity condition (8)
and (9) for the input bipartite state.

Now we want to infer the classicality of the output separable
state, and for that we use a standard classicality measure, which
is P-representability \cite{schur}: A state \be\rho=\int d\alpha^2
d\beta^2
P(\alpha,\beta)|\alpha,\beta\rangle\langle\alpha,\beta|\ee is
P-representable if $P(\alpha,\beta)$ is a positive Glauber
P-function less (or equally) singular than the delta distribution.
This is possible \cite{englert,schur} whenever ${\bf V^\prime
}-\frac 12 {\bf I}\ge
 0$, which by consequence implies that both output parties must be
 P-representable themselves, $ {\bf V_1^\prime }-\frac 12 {\bf I}\ge
 0$, ${\bf V_2^\prime }-\frac 12 {\bf I}\ge
 0$, or in terms of the observable quantities
\be n_1^\prime\ge|m_1^\prime|+\frac 1
2,\;\;\;n_2^\prime\ge|m_2^\prime|+\frac 1 2,\label{sept}\ee
respectively. Matter-of-factly  either conditions (\ref{sept})
when explicitly written in terms of the input parameters
correspond exactly to the separability condition (\ref{sep}) and
(\ref{sep2}). Thus a measure of the classicality
(P-representability) of any of the output parties represents,
one-to-one, a measure of the input state separability condition.

\section{Quantifying entanglement through one-party classicality measure}

We are now in position to infer about a quantitative measure of
the nature of nonseparable states through the properties of one of
the ${\bf M}$-transformed one-party output state.  Observe that to
any Gaussian nonclassical output state, satisfying
$\sqrt{|m_i^\prime|^2+\frac 1 4}\le n_i^\prime<|m_i^\prime|+ \frac
1 2$, $i=1$ or $2$, there is a one-to-one corresponding input
bipartite mixed entangled Gaussian state belonging to the SSLD. An
immediate consequence is that {any available quantitative Gaussian
measure (a measure that preserves the Gaussian structure of the
state) of nonclassicality of one of the one-party output states
can be regarded as a quantitative Gaussian measure  for the degree
of entanglement of the bipartite input state}. The veracity of
this statement rests on the following proposition:

   { Proposition 1:} {\it Let $\rho$ be a bipartite input Gaussian state of the SSLD and
    $\rho^\prime_1\otimes\rho^\prime_2$ a ${\bf M}$-transformed separable output
    state. In terms of Gaussian measures,
    the less classical are the ${\bf M}$-transformed output states,
    $\rho^\prime_1$ or
    $\rho^\prime_2$,
  the more entangled is $\rho$.}


  To prove proposition 1, an important relation between bipartite
   and one-party state Gaussian measures will be derived.

{ Lemma 1:} Let ${\mathcal{A}}=Tr[f(A,\rho)]$ be the effect of the
measure $f(A,\cdot)$ over the Gaussian state $\rho$, being
$f(A,\cdot)$ an analytic function of a trace class Gaussian
operation $A$ and $\rho$. Any nonlocal Gaussian measure
$f(A,\cdot)$ of a bipartite Gaussian state $\rho_{in}$ can be
converted into local Gaussian measures $
{\mathcal{A}}=Tr_1[f(A_{out}^{(1)},\rho_{out}^{(1)})]\;Tr_2[f(A_{out}^{(2)}\,
\rho_{out}^{(2)})] $ of the two one-party output states by
nonlocal rotations $\bf M$.

{ Lemma 2:} Any local Gaussian measure of a bipartite separable
Gaussian state: $Tr_1[f(A_{out}^{(1)},\cdot)]$, or
$Tr_2[f(A_{out}^{(2)}\,\cdot)] $,
 can be converted into a nonlocal Gaussian
measure of a nonseparable bipartite Gaussian state under a
reduction and an appropriate nonlocal rotation $\bf M^{-1}$.

{\it Proof of Lemma 1}: Since unitary transformations do not alter
the effect content it is immediate that ${\mathcal{A}}=Tr[f(B
AB^\dagger,B\rho_{in}B^\dagger)]$. Let $\mathbf{\Gamma}$ be the
Gaussian characteristic function associated to $A$ \cite{giedke}
such that $\mathbf{M}$ acts on $\mathbf{\Gamma}$ and $\mathbf{V}$
in similar
 fashion (Eqs. (16-18)). Thus ${\mathcal{A}}=Tr[f(A_{out},
 \rho_{out})]$, where $A_{out}$ is
the output Gaussian operation with covariance matrix
$\mathbf{\Gamma^\prime=M^{-1}\Gamma
 M}$. With the conditions (21) and (22) to
bring $\mathbf{V^\prime}$ and $\mathbf{\Gamma^\prime}$ to the
(separable) block-diagonal form satisfied, it is immediate
that\br\nonumber
\label{eq1}{\mathcal{A}}=Tr_1[f(A_{out}^{(1)},\rho_{out}^{(1)})]\;Tr_2[f(A_{out}^{(2)}\,
\rho_{out}^{(2)})].\hspace{2cm}\blacksquare\er

{\it Proof of Lemma 2}: The explicit Gaussian forms of
$f(A_{out}^{(i)},\cdot)$ and $A_{out}^{(i)}$ itself, are only
given when the output measurement is specified, but Lemma 1 must
hold even for the situation where the measurement in either of the
output ports is operationally equivalent to a reduction:
$Tr_2[f(A_{out}^{(2)},\rho_{out}^{(2)})]\equiv 1$ or
$Tr_1[f(A_{out}^{(1)},\rho_{out}^{(1)})]\equiv 1$, exclusively,
such that
\br\label{eq2}{\mathcal{A}}=Tr_i[f(A_{out}^{(i)},\rho_{out}^{(i)})],
\, (i=1, \mbox{ or }2), \nonumber\er which by definition is equal
to $Tr[f(A,\rho_{in})]$. The input effect can be computed through
the corresponding characteristic function of
$f(A_{out}^{(i)},\cdot)$, which by hypothesis is Gaussian, and the
proof is completed in similar fashion to the proof of Lemma
1.\hspace{5cm}$\blacksquare$

{\it Proof of Proposition 1}: From Lemma 1 and 2, it is immediate
that an entanglement quantification via a one-party
nonclassicality measure is possible, whenever both the output and
input measures are Gaussian and correspond, respectively, to
nonclassicality  and entanglement quantifications. An important
measure falling inside this description is the Uhlmann Fidelity
\cite{bures} between any two Gaussian states $\rho$ and $\sigma$:
${\cal
F}(\rho,\sigma)=\{Tr[(\sqrt{\rho}\sigma\sqrt{\rho})^{1/2}]\}^2$,
which is central for the calculation of the Bures distance
\cite{bures}: 
$d_B(\rho,\sigma)=(2-2\sqrt{\cal F}(\rho,\sigma))$.
  The Bures distance
was recently identified as a quantification for nonclassicality
\cite{paulina1} and entanglement \cite{vedral,paulina2}, for one
and two parties states, respectively, when $d_B(\rho,\sigma)$ is
minimized over the possible referential $\sigma$ belonging to the
Gaussian subset of states. In fact from  and Lemma 1 it is
immediate that ${\cal F}(\rho_{in},\sigma_{in})={\cal
F}(\rho^{(1)}_{out},\sigma^{(1)}_{out}){\cal
F}(\rho^{(2)}_{out},\sigma^{(2)}_{out})$ and
the input-output Bures metrics are related as
\br\label{bures2}
d_B(\rho_{in},\sigma_{in})&=&d_B(\rho^{(1)}_{out},\sigma^{(1)}_{out})+d_B(\rho^{(2)}_{out},\sigma^{(2)}_{out})\nonumber\\
&&-\frac 1 2
d_B(\rho^{(1)}_{out},\sigma^{(1)}_{out})d_B(\rho^{(2)}_{out},\sigma^{(2)}_{out}).\er
 Inversely, if the measurement is made on the port $i$ only, meaning that
  $\sigma^{j}_{out}=\rho^{j}_{out}$, $j=i-(-1)^i$, ($i=1$, $2$), then
from Lemma 2, ${\cal F}(\rho^{(i)}_{out},\sigma^{(i)}_{out})={\cal
F}(\rho_{in},\sigma_{in})$, and \be\label{bures3}
d_B(\rho^{(i)}_{out},\sigma^{(i)}_{out})=d_B(\rho_{in},\sigma_{in}),\ee
Here the corresponding $\sigma_{in}$ must necessarily be
   $B^\dagger \sigma^{(i)}_{out}\,\rho^{(j)}_{out}B $.
   Both situations reflect the direct
connection between two measures of classicality (distance from a
reference classical state \cite{paulina2}) and entanglement
(distance from a reference separable state \cite{paulina1}), and
thus the observance of Proposition 1.\hspace{4.8cm}$\blacksquare$

\section{Example}
Next we illustrate the above discussion for the
TMTSS. To simplify  our  calculations, instead of using the Bures
distance directly, we
 introduce a simple pictorial entanglement
measure ${\cal
E}(\rho,\sigma)=1-d_B(\rho,\sigma)/d_B(\rho_{sep},\sigma)$ (see
\cite{paulina2}), where
 $\sigma$ is a maximally entangled (pure) Gaussian state
 represented by the equality in (\ref{pos}). $\rho_{sep}$ is the separable
density operator, introduced to give ${\cal E}(\rho,\sigma)=0$  at
the  separability boundary, and obtained simply by tracing out one
of the parties in $\sigma$. Notice that a $\sigma$ pure does not
allows that ${\cal E}(\rho,\sigma)$ be an entanglement monotone
(see \cite{vedral,paulina2}), once it can always be increased by
appropriate local operations on $\rho$. But this choice suffices
for our illustrative purpose, simplifying the calculations, since
the Uhlmann fidelity is simply given by ${\cal
F}(\rho,\sigma)=Tr(\rho\sigma)$. As such for one party systems the
nonclassicality measure is related to the distance from a pure
nonclassical one-mode squeezed state,
$\sigma^{(i)}_{out}=|\psi_i\rangle\langle\psi_i|:$\be
|\psi_i\rangle
=e^{-r({a^\dagger}^2-a^2)}|0\rangle=\sqrt{1-\lambda^2}\sum_{k=0}^\infty\lambda^k|k\rangle,\ee
while for bipartite systems the entanglement measure is related to
the distance from a pure two-mode squeezed state, $
\label{ent}\sigma=|\Psi\rangle\langle\Psi|:$\be|\Psi\rangle =
e^{-r(a^\dagger
b^\dagger-ab)}|0,0\rangle=\sqrt{1-\lambda^2}\sum_{k=0}^\infty\lambda^k|k,k\rangle,\ee
both in absence of noise and with $\lambda=\tanh r$, $n=\cosh
2r/2$, $m=-\sinh 2r/2$, and $r$  the squeezing parameter. Those
states maximize our pictorial measures (minimize the Bures
distance) for $\lambda\rightarrow \infty$ corresponding to
irreducible maximal nonclassical and entangled states,
respectively.
A nonseparable TMTSS generated in a nonlinear crystal \cite{daffer} is bounded by 
$ m+\frac 1 2>n\ge \sqrt{m^2+\frac 1 4}$.
Now if the two parametric-down-converted beams are mixed at an
ideal (lossless) $50:50$ beam-splitter ($\theta=\pi/4$), the
output port nonclassicality condition of either beams reads
$\label{fim} n^\prime< m^\prime+\frac 1 2$, which converted into
the input parameters corresponds to the input state
nonseparability bound.
The output $i$-port Uhlmann fidelity is given by \be{\cal
F}(\rho^{(i)}_{out},\sigma^{(i)}_{out})=\left[{n_i^\prime}^2-{m_i^\prime}^2+n_i^\prime
\cosh 2r+m_i^\prime \sinh 2r+\frac 1 4\right]^{-1/2},\ee  which is
precisely the Fidelity  between the TMTSS and the pure two mode
squeezed state (without primes). Both the separability boundary
and ${\cal E}(\rho,\sigma)$ of the two-parametric-down-converted
fields are simply given by output local classicality measures
based on the Bures distance, which can be achieved by
 homodyning one of the
output ports to a pure squeezed state of reference, as fully
described in \cite{kimmunro}. In Fig. 1 we plot ${\cal
E}(\rho,\sigma)$ in the $\{n,m\}$ parameter space for $r=1$,
together with the separability limiting bound (dashed line), as
given by the corresponding nonclassicality measure. Notice that
although pictorial and simplified the introduced measure is also
able to quantify the amount of entanglement in pure states in
relation to the maximally entangled pure state, obtained for
$r\rightarrow \infty$.


\section{Conclusion}
We have given a simple direct relation between entanglement
measures for the special class of Gaussian input quantum states
belonging to the SSLD in terms of the output states
nonclassicality. Under nonlocal rotations both the separability
boundary and quantification of the presented amount of
entanglement are equivalent to one-party classicality
(P-representability) boundary and to a nonclassicality
quantification, respectively. It would be certainly interesting to
analyze the amount of information about entanglement that can be
obtained by appropriate inverse transformation of any other
available one-party classicality criteria (such as the
hierarchical measure of classicality \cite{vogel}). Our
preliminary results show that the so established inequalities are
weaker than the above separability ones, in the same sense that
any other classicality criterion is weaker than the
P-representability one. This point is left for further
investigation.



\end{multicols}
\newpage
{\bf Figure Captions}

\begin{figure}
 \centerline{$\;$\hskip 0truecm\psfig{figure=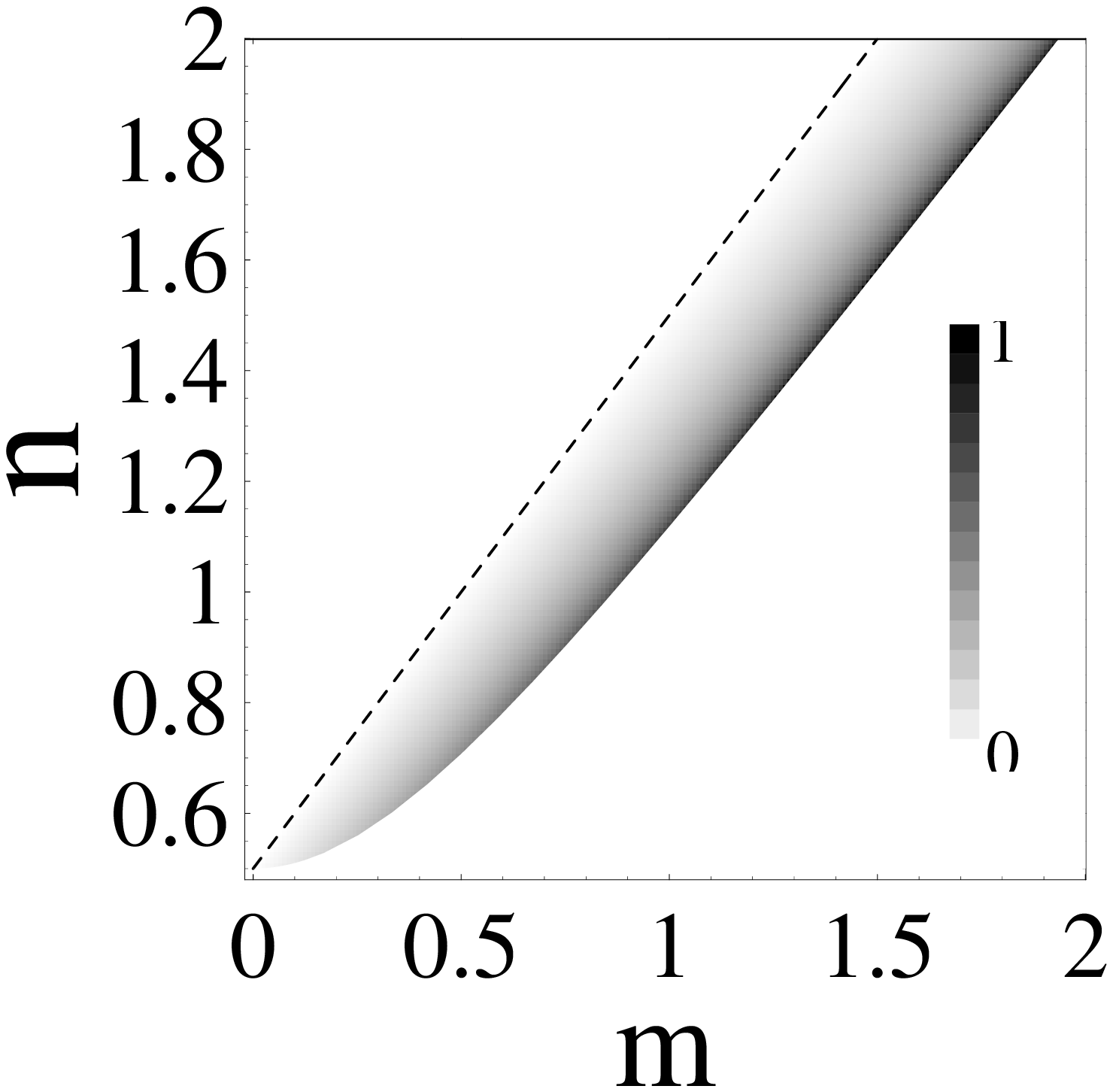,height=10cm}}
\parbox{8cm}{\small Fig.1. Degree of entanglement ${\cal E}(\rho,\sigma)$ for ($r=1$)- TMTSS.} \label{fig1}
\end{figure}
\end{document}